\documentclass[aps,pre,twocolumn,superscriptaddress]{revtex4}  
\usepackage{graphicx,epsfig}  
\usepackage{dcolumn}   
\usepackage{bm}        
\usepackage{amssymb}   
\usepackage{color}
\usepackage{mathtools}
\usepackage{esvect}
\usepackage{float}

\begin{document}

\title{ Harmonic generation in the interaction of laser with a magnetized overdense plasma }

\author{Srimanta Maity}
\email {srimantamaity96@gmail.com}
\affiliation{Department of Physics, Indian Institute of Technology Delhi, Hauz Khas, New Delhi 110016, India}

\author{Devshree Mandal}
\affiliation{Institute for Plasma Research, HBNI, Bhat, Gandhinagar 382428, India}
\affiliation{Homi Bhabha National Institute, Mumbai, 400094, India}
\author{Ayushi Vashistha}
\affiliation{Institute for Plasma Research, HBNI, Bhat, Gandhinagar 382428, India}
\affiliation{Homi Bhabha National Institute, Mumbai, 400094, India}
\author{Laxman Prasad Goswami}
\affiliation{Department of Physics, Indian Institute of Technology Delhi, Hauz Khas, New Delhi 110016, India}

\author{Amita Das}
\affiliation{Department of Physics, Indian Institute of Technology Delhi, Hauz Khas, New Delhi 110016, India}

\begin{abstract}
The mechanism of harmonic generation in both O and X-mode configurations for a magnetized plasma has been explored here in detail with the help of Particle-In-Cell (PIC) simulations. A detailed characterization of both the reflected and transmitted electromagnetic radiation propagating in the bulk of the plasma has been carried out for this purpose. The efficiency of harmonic generation is shown to increase with the incident laser intensity.
   Dependency of harmonic efficiency has also been found on magnetic field strength. This work demonstrates that there is an optimum value of the magnetic field at which the efficiency of harmonic generation maximizes. The observations are  in agreement with theoretical analysis. For O-mode configuration, this is compelling as the harmonic generation provides 
 for a mechanism by which laser energy can propagate inside an overdense plasma region.   

\end{abstract}

\maketitle

\section{Introduction}
 \label{intro}
 Laser-plasma interaction has been an important field attracting research interest for several decades \cite{liu2019high, kaw2017nonlinear}. The research has been focused on many exciting areas of physics, e.g., long scale magnetic field generation \cite{bret2009weibel, das2020boundary}, nonlinear electromagnetic structure formation \cite{verma2016stability, mandal2020spontaneous, yadav2021nonlinear}, wave breaking \cite{modena1995electron, bera2021effect}, particle acceleration \cite{tajima1979laser, modena1995electron, faure2004laser}, x-ray source \cite{rousse2004production, corde2013femtosecond} and gamma-ray source \cite{cipiccia2011gamma}, etc. Traditionally, laser-plasma interaction study has essentially been focused and limited to the regime of unmagnetized plasma response. This has been so as the required external magnetic field to elicit a magnetized response from plasma medium at the laser frequency is quite high and not possible to achieve in the laboratory. Lately, there have been technological developments in this direction, and magnetic fields of the order of kilo Tesla has been achieved \cite{nakamura2018record}.

 The regime of laser interacting with a magnetized plasma has thus caught attention lately. Studies investigating new possibilities of laser energy absorption, penetration, and characteristics mode propagation inside the plasma, have been carried out in detail \cite{Kumar_2019, Vashistha_2020, ayushi_nf}. These studies have used the X-mode geometry for which the laser electric field is normal to the applied external magnetic field. The O-mode configuration, however, has not been studied so far. It is generally believed that such a configuration will exhibit a similar response as that of an unmagnetized plasma and would have nothing new to offer. With the help of Particle-In-Cell (PIC) simulation, we have shown, that the  O-mode geometry has surprises to offer.  While the laser energy cannot penetrate an unmagnetized overdense plasma, in the case of O-mode configuration, this happens with the help of harmonic generation (HG) at the laser-plasma interface.  
 It is shown that a  part of laser energy gets converted into higher harmonics in the presence of an externally applied magnetic field and can propagate inside the plasma if the plasma is underdense at this high frequency.

The physics of harmonic generation of electromagnetic (EM) radiation is itself an important area of investigation \cite{margenau1948theory, sodha1970theory}. It has been studied in the context of laser-plasma interaction for several decades now \cite{burnett1977harmonic, carman1981observation, Rev_harmonic, ganeev2012high}. The high-harmonic observations in laboratory plasmas have opened up a wide range of applications. It is considered as one of the most efficient techniques known to obtain electromagnetic waves of higher frequency in a controlled manner \cite{tsakiris2006route, dromey2006high}. HG has been used as a distant probe for the detection of turbulence in a toroidal magnetically confined plasma \cite{ajendouz2007high}. Polarization measurement of the harmonic has also been used to detect the poloidal magnetic field profile in Tokamak devices \cite{cano1975poloidal}. Recently, the second harmonic radiation generated in the interaction of laser beams with an underdense plasma has been used to experimentally verify some of the fundamental properties of photons, including the conservation of total angular momentum \cite{huang2020conservation}.

In the review article by \cite{Rev_harmonic}, a nonlinear fluid model has been discussed to predict the harmonic generation in the reflected radiation from an overdense plasma surface in the absence of an external magnetic field. An equivalent model was also formulated by 
\cite{bulanov1994interaction} and \cite{lichters1996short} predicting the existence of so-called selection rules for the polarization of reflected harmonics from an overdense plasma surface. As already stated, we show here, with the help of PIC simulations, that a magnetized plasma provides another mechanism of harmonic generation. High-order harmonic generation in the interaction of laser with a magnetized plasma has been reported in several previous studies. The second
harmonic generation in a uniform magnetized plasma has been studied by \cite{jha2007second}. Second and third-harmonic generation in the interaction of laser fields with a magnetized plasma having a density below the critical density has been reported by \cite{ghorbanalilu2012second}. Second harmonics generation in the reflected and transmitted radiation by an obliquely p-polarized laser pulse propagating through a homogeneous, underdense, and transversely magnetized plasma is studied by \cite{ghorbanalilu2017reflected}. The second harmonic generation of a relativistic chirped laser pulse propagating through homogeneous magnetized plasma is studied by \cite{kant2016second}. These studies are mostly analytical, involving approximations. In all these studies, plasma was assumed to be homogeneous and underdense so that a laser could propagate through the plasma. In the present study, we have performed PIC simulations considering a finite laser pulse (O-mode configuration) incident on an overdense magnetized plasma surface where the original laser pulse cannot penetrate inside the bulk plasma. It only interacts with the plasma species at the vacuum-plasma interface.

 In the PIC study reported by \cite{mu2016effect}, they observed second harmonic generation in the reflected radiation from a solid, dense plasma surface in the presence of an external magnetic field. They explored the harmonic generation efficiency with the variation of external magnetic field and pre-plasma scale length. In the present study, we have observed the presence of higher harmonics in the reflected as well as transmitted radiations. Here, we have mainly concentrated on the characterization of the harmonic radiations transmitted inside the bulk plasma, which was unexplored in the previous study. The harmonics get generated at the vacuum plasma interface and propagate inside the plasma as well as in the vacuum region.  The conditions for channeling the harmonic radiations generated at the vacuum plasma interface inside the bulk plasma have been identified and analyzed. It occurs in both O and X-mode configurations. We have provided a comparison and contrasted both configurations for a harmonic generation. Simulation observations for the case with and without externally applied magnetic fields have been compared and discussed. The conversion efficiency of HG for a wide range of external magnetic fields has been analyzed. We consider the laser to be incident normal to the plasma surface for various laser and plasma parameters and have observed conversion efficiency as high as $1.77\% $ for the second harmonic. The conversion efficiency for these higher harmonics both in the transmitted (inside the plasma) and reflected (in vacuum) radiations has been shown to depend on the strength of the laser and external magnetic field. 
Furthermore, we have also characterized the following properties of harmonic radiation in detail: (i) the dispersion properties of the observed higher harmonics, (ii) polarization of the higher harmonic radiation, and (iii) forbidden frequencies for given plasma and EM wave parameters.

  The paper has been organized as follows. In section 2, we describe the simulation setup.  Section 3 contains 
  the observations. The various subsections therein describe the generation and characterization of harmonics. In section 4, we provide a summary and conclusion. We have provided an approximated analytical calculation for the mechanism of HG in the appendix.


\section{Simulation Details}
\label{picsim}

\begin{figure*}
\centering
   \includegraphics[height = 7.0cm,width = 13.5cm]{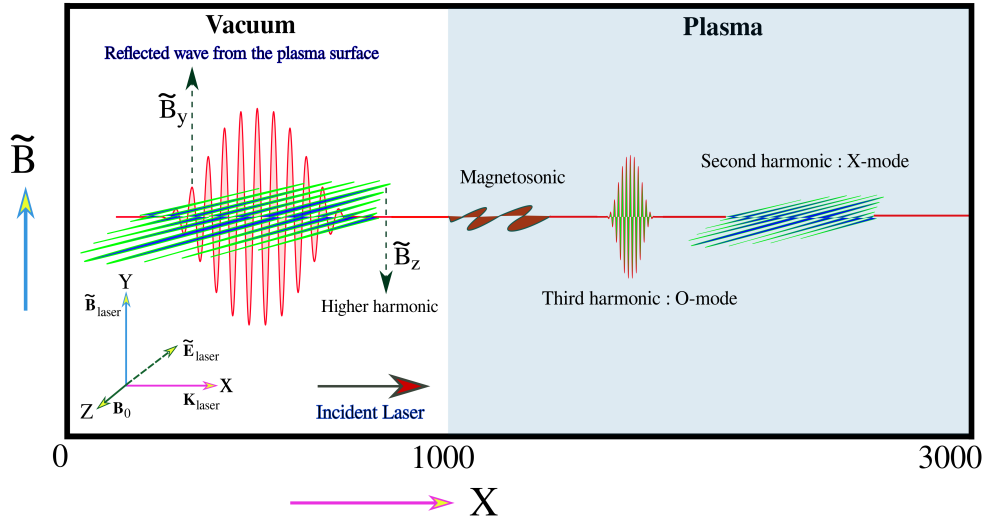}
   
   \caption{A summary of the observations of this study has been shown in this schematic. We have performed 1D PIC simulation (along $\hat x$) with a laser being incident on the plasma surface at $x = 1000$. The external magnetic field $B_0$ has been applied along $z-$direction. The polarization of the incident laser has been chosen in O-mode configuration in this schematic, i.e., the electric field of the incident laser pulse is oscillating along the direction of the external magnetic field $B_0$. As the laser interacts with the plasma surface, it generates higher harmonics with different polarization in the reflected and transmitted radiations, as shown in the schematic. Magnetosonic disturbance has also been observed in these interactions.}

  \label{schmtc}
\end{figure*}

 We have carried out one-dimensional Particle-In-Cell Simulations using OSIRIS 4.0 framework \cite{Fonseca2002,osiris,Hemker} for our study. Our simulation geometry has been shown in Fig. \ref{schmtc}. It has a longitudinal extent of $3000 d_{e} $ with plasma boundary starting from  $x=1000d_{e}$. Here, $d_e$ is the electron skin depth $c/\omega_{pe}$ and $c$ is the speed of light in vacuum. We have  chosen $60000$ grid points, which corresponds to  $dx= 0.05d_e$. The  number of particles per cell has been chosen to be $8$. Time has been normalized by $t_N=\omega_{pe}^{-1}$, where $\omega_{pe} $ is the plasma frequency corresponding to the density $n_0$. The length is normalized by $ x_N= c/\omega_{pe}= d_e$, and fields by $B_N=E_N= m_ec\omega_{pe}/e$, where $m_e$ and  $e$ represent the rest mass of an electron and the magnitude of an electronic charge, respectively. The external magnetic ($B_0=2.5$ in normalized units) field has been applied along the $\hat{z}$ direction.

 Table \ref{simulation} presents laser and plasma parameters in normalized units. A possible set of values in the standard unit has also been provided. We have treated both electron and ion dynamics in our simulations with a reduced mass of ions $m_i=100m_e$. For some cases, we have also treated the ions to be infinitely massive and acting merely as a stationary background. The inferences on HG were not altered for the two cases. A laser pulse with intensity $\approx 3 \times 10^{15} $ $W/cm^2$ (for normalized vector potential, $a_0=eE_l/m\omega_lc=0.5$) and frequency $\omega_{l}= 0.4 \omega_{pe}$ is  incident normally from the left side of plasma. The electric field of the laser $E_l$ is chosen to be along $\hat{z}$ for the O-mode configuration and along $\hat{y}$ for the X-mode configuration. In Fig. \ref{schmtc} the O-mode configuration has been depicted in a schematic representation. The longitudinal profile of laser pulse is a polynomial function with rise and fall time of $100 \omega_{pe}^{-1}$ which translates to $200fs$ and it starts from $x=950 d_{e}$.  The value of the  external magnetic field  is chosen  to elicit magnetized response of electrons while ions remain unmagnetized at the laser frequency  i.e. $\omega_{ce}> \omega_{l}> \omega_{ci}$, where  $\omega_{ci}, \omega_{ce}$ are cyclotron frequencies of ion and electron respectively while $\omega_{l}$ is laser frequency. Absorbing boundary conditions are used for fields and particles. For an unmagnetized plasma (i.e., in the absence of an external magnetic field), the laser cannot penetrate inside the plasma. In the X-mode configuration, even in the overdense case,  if the laser frequency lies in the passband of magnetized plasma, it propagates inside and generates lower hybrid and magnetosonic excitation, as has been illustrated in some of the earlier works \cite{Kumar_2019, Vashistha_2020, ayushi_nf}. However, for  O-mode configuration, the laser cannot propagate inside plasma if it is overdense. In this case, the laser light only penetrates the plasma up to skin depth order. We observe that this is sufficient for the generation of harmonics in the O-mode configuration. For the higher frequency of harmonics that get generated, the plasma becomes underdense. In such a situation, the generated harmonic radiation is free to propagate inside the plasma. 

In our PIC simulation study, we have not included Coulomb collisions. The quiver velocity of electrons 
would be high at the laser intensities of $\sim 10^{15}$ $W/cm^2$ considered in our study.  We, therefore,  feel that the collisions will not have any significant role to play \cite{dendy1995plasma}. In our study, the initial temperature of electrons are assumed to be very low ($T_e = 0.05$ $eV$). Such an assumption is valid because the temperature remains small compared to the typical oscillation energy of electrons in intense laser fields \cite{gibbon2005short}. Thus, the plasma temperature will not affect the HG mechanism presented here. We have done a comparative study of temperature effect on the generation of harmonics too (in section \ref{mechansm}) which illustrates the above point.

\begin{table}
\centering
\caption{Simulation parameters: In normalized units and possible values in standard units.} 
\vspace{0.2cm}
	\begin{tabular}{|p{2.5cm}||p{2.5cm}||p{2.5cm}|}
		
		\hline
		\textcolor{black}{Parameters}& \textcolor{black}{Normalized Value}& 	\textcolor{black}{A possible value in standard unit}\\
		
		\hline
		\hline
		\multicolumn{3}{|c|}{\textcolor{black}{Laser Parameters}} \\
		\hline
		Frequency ($\omega_l$)&\centering$ 0.4 \omega_{pe}$& $1.78 \times 10^{14} $rad/s\\
		
		\hline
		Wavelength&\centering$15.7c/\omega_{pe}$ &$10.6\mu m$\\
		\hline
		Intensity&\centering$a_{0} =0.5$&$ 3.04 \times10^{15} W/cm^2$ \\
		\hline
		\multicolumn{3}{|c|}{\textcolor{black}{Plasma Parameters}} \\
		\hline
		Number density($n_0$)&\centering$1$&$6.2 \times 10^{19}$ $ cm^{-3}$\\
		\hline
		Electron Plasma frequency ($\omega_{pe}$)& \centering$1$ &$4.44 \times 10^{14} $rad/s\\
		\hline
		Electron skin depth ($c/\omega_{pe}$)& \centering$1$ & $0.68\mu m$ \\
	\hline
		\multicolumn{3}{|c|}{\textcolor{black}{External Fields }} \\
		\hline
	Magnetic Field ($B_0$) &	\centering$2.5$ & $\approx$ 6.3 kT \\
		\hline
	\end{tabular}
		\label{simulation}

\end{table}

\section{Observations and Discussion}
It is well known that in the X-mode configuration, the EM radiation of the laser penetrates the plasma in the respective permitted passbands. In this case, bulk plasma can interact with the incident radiation although $\omega_l < \omega_{pe}$.  In the O-mode, however, the dispersion relation being identical to the unmagnetized case, there is no propagation when the EM wave frequency is smaller than the plasma frequency.  The laser-plasma interaction, in this case, is thus confined only within the electron skin depth layer. The plasma within the skin depth responds to the Lorentz force acted upon by the electric and magnetic field of the EM wave and the applied external magnetic field. The incident EM radiation has a finite spatial pulse profile in the longitudinal direction. 
 The finite spatial profile of the laser pulse provides for an additional ponderomotive force to the plasma medium. We have chosen to work in the frequency domain ( shown in the table \ref{simulation}) for which 
  the condition $\omega_{ci}<\omega_l<\omega_{ce}$ is satisfied. Here, $\omega_l$ defines the laser frequency, and $\omega_{ce}$ and $\omega_{ci}$ represent the electron and ion cyclotron frequency, respectively. Thus, electrons would have a magnetized response to offer in the time period corresponding to a  laser cycle, whereas ions would be unmagnetized. 
 We now present various features of our observations in the following subsections. 

\subsection{HG in O-mode configuration $(\protect\vv{\bm{E}}_l \parallel \protect\vv{\bm{B}}_0)$} \label{hrmnics_Omode}

\begin{figure*}
\centering
   \includegraphics[height = 3.5cm,width = 13.5cm]{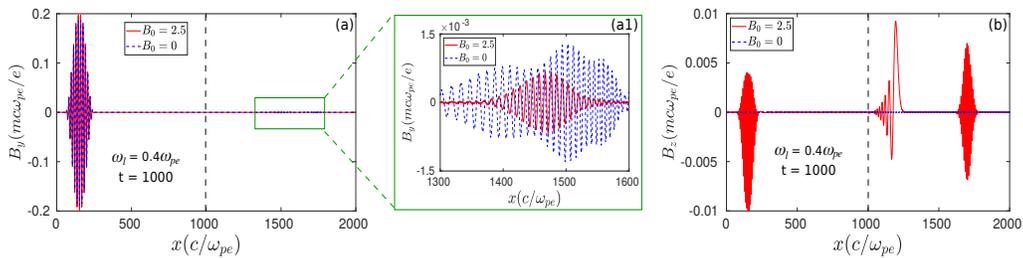}
   
   \caption{Transverse time varying magnetic fields $B_y$ and $B_z$ with respect to $x$ have been shown at a particular instant of time $t = 1000$ (when laser already gets reflected back from the system) in subplots (a) and (b), respectively. In subplot (a1), $B_y$, which exists inside the plasma, has been shown on a different scale. Here, the black dotted line at $x = 1000$ represents the plasma surface. It is to be noted that the electromagnetic fields $\vec{B_l}$ and $\vec{E_l}$ 
   of the incident laser pulse were along $\hat y$ and $\hat z$ directions, respectively. In this figure, red lines represent for $B_0 = 2.5$ with $E_l \parallel B_0$ and blue dotted lines are for $B_0 = 0$.}

  \label{field_0p4}
\end{figure*}


\begin{figure*}
\centering
   \includegraphics[height = 10.0cm,width = 13.0cm]{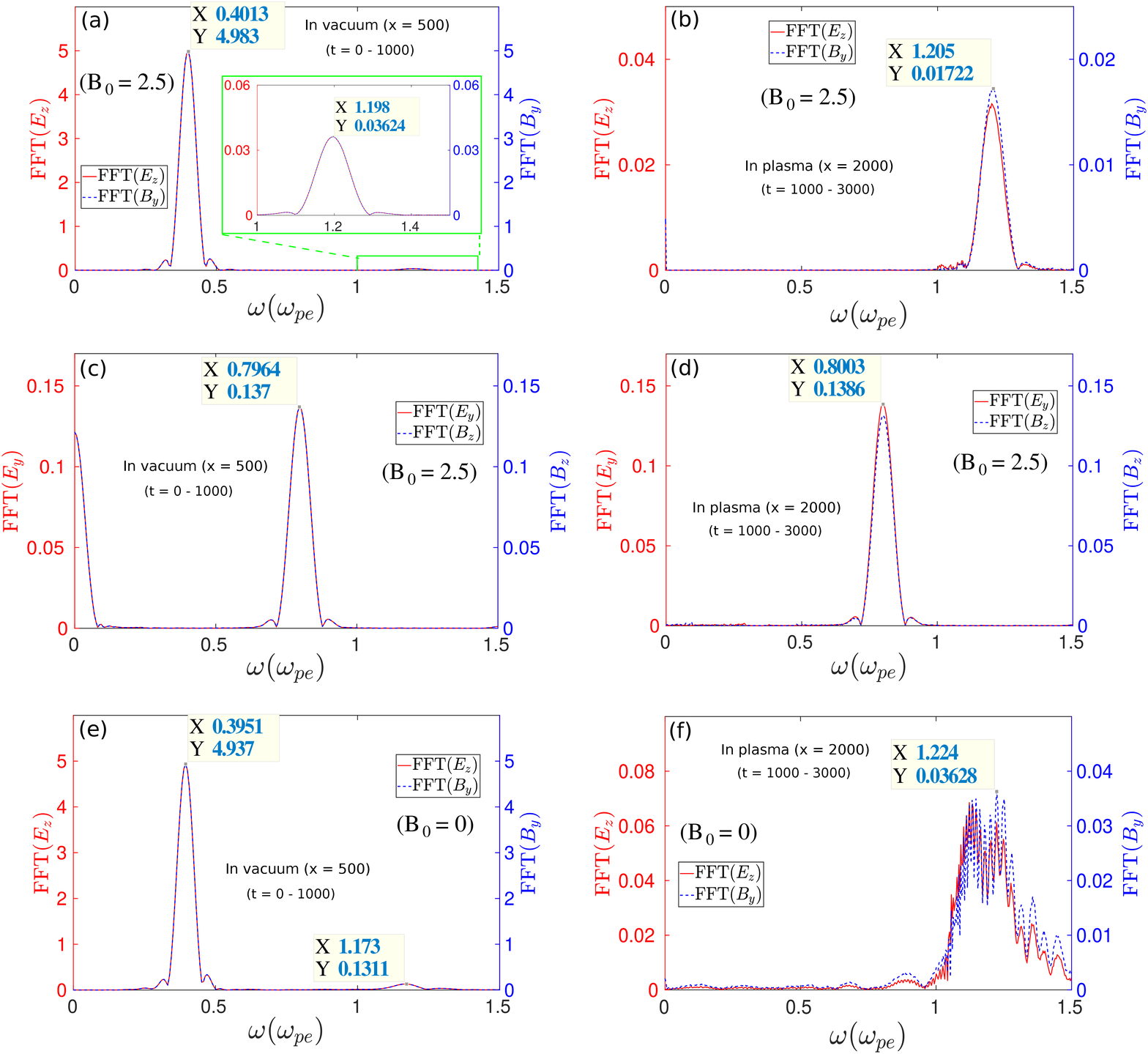}
   
   \caption{Fourier transform of electromagnetic fields with time after laser reflected from the plasma surface. In subplot (a) and (b), the FFT of $E_z$ and $B_y$ with time in vacuum ($x = 500$) and the bulk plasma ($x = 2000$) have been shown, respectively. The same has been shown for the fields ($E_y$, $B_z$) in subplots (c) and (d), respectively. In the subplots (a)-(d), the external magnetic field ($B_0$) is considered to be $2.5$. These FFTs clearly indicate that the higher harmonics have been generated, and they are present in both vacuum and the bulk plasma. Subplots (e) and (f) represent the time FFT of $E_z$ and $B_y$ without any external magnetic field ($B_0 = 0$) in vacuum and plasma, respectively.
}

  \label{fft_0p4}
\end{figure*}

We first consider the case when the frequency of the incident laser pulse was chosen to be $0.4 \omega_{pe}$ and the polarization of the laser fields were considered to be in O-mode configuration, i.e., $\vv{\bm{E}}_l \parallel \vv{\bm{B}}_0$ (in $\hat z$ direction). Here, $\bm{E}_l$ is the laser electric field, and $\bm{B}_0$ is the externally applied magnetic field. The transverse $\hat y$ and $\hat z$ components of the magnetic field, $B_y$ and $B_z$ have been shown by red solid line for $B_0 = 2.5$ at a particular instant of time $t = 1000$ in subplots (a) and (b) of Fig. \ref{field_0p4}, respectively. It is to be noted that at time $t = 0$, laser pulse with the electromagnetic fields $B_y$ and $E_z$ was set to propagate along positive $\hat x$ direction from the location $x = 950$. Thus, the structure in $B_y$ present in the vacuum region ($x \approx 200$) at $t = 1000$, as seen from subplot (a), is associated with the reflected part of the incident laser propagating along $-\hat x$ direction. A small fraction of $B_y$ is also present inside the bulk plasma, as depicted in the zoomed scale in subplot (a1). We will identify this structure as the third harmonic radiation (and the higher odd harmonics) in the consecutive section. In the subplot (b) of Fig. \ref{field_0p4}, it is seen that the $\hat z-$component of the oscillating magnetic field $B_z$ which was not present before the laser hits the plasma surface, has been produced at a later time and exists in both vacuum and bulk plasma. There are two types of disturbances observed inside the bulk plasma, as can be seen from the subplot (b). One is the large-scale disturbance near the plasma surface, which will be identified as the magnetosonic perturbation. And another disturbance moving with the faster group velocity. In the following discussion, we will show that this is the second harmonic radiation (even harmonics) generated due to the interaction of laser pulse with the plasma particles. The simulation observations for $B_0 = 0$ have also been depicted in Fig. \ref{field_0p4} by blue dotted lines. It is seen that in this case, also the $\hat y$ component of the magnetic field, i.e., $B_y$  is present in both reflected and transmitted radiations. However, no $\hat z$ component of the magnetic field, i.e., $B_z$, is generated in this case, as can be seen in subplot (b) of Fig. \ref{field_0p4}.

 The time FFT (Fast Fourier Transform) of these reflected and transmitted radiations has been shown in Fig. \ref{fft_0p4}. The FFTs of $E_z$ and $B_y$ at the location $x = 500$ (vacuum)  show  two distinct peaks at the frequency $\omega \approx 0.4\omega_{pe}$ and $\omega \approx 1.2\omega_{pe}$ (subplot (a) of Fig. \ref{fft_0p4}). It is important to recall that at $t = 0$, the incident laser pulse was located in between $x= 750$ and $950$. The first peak with higher power at the location $\omega \approx 0.4\omega_{pe}$ is essentially the original laser pulse which has been reflected from the plasma surface, as plasma is overdense. The second peak located at $\omega \approx 1.2\omega_{pe}$ is the third harmonic radiation. The third harmonic is also present inside the bulk plasma and has been demonstrated by carrying out the  FFT in time for the  $E_z$ and $B_y$ signals at the location $x = 2000$ (plasma), as has been shown in subplot (b) of Fig. \ref{fft_0p4}. Thus, it is now clear that the small disturbance in $B_y$ present inside the plasma, as shown in subplot (a1) of Fig. \ref{field_0p4} is essentially associated with the third harmonic radiation. The  FFT of $E_y$ and $B_z$ in time at the locations $x = 500$ and $2000$ have been shown in subplots (c) and (d) of Fig. \ref{fft_0p4}, respectively. The FFT at the location $x = 2000$ (plasma), in subplot (d), has been evaluated within a time window $t = 1000$ to $3000$. This is to eliminate the slowly moving magnetosonic disturbance as shown in the subplot (b) of Fig. \ref{field_0p4}. It is seen that the frequency spectrum of $E_y$ and $B_z$ has a distinct peak at the location $\omega \approx 0.8\omega_{pe}$ in both cases (vacuum and plasma). This ensures that the second harmonic radiation has been generated and travels in both vacuum and plasma mediums. The observation of second and third harmonics in the reflected radiation in the presence of a transverse external magnetic field had been reported in an earlier study by \cite{mu2016effect}. However, in our study, we have observed the harmonic radiations in both reflected and transmitted radiations, as has been shown in Fig. \ref{fft_0p4}. In the case without an external magnetic field, the third harmonic (odd harmonics) shows up with the polarization ($E_z$, $B_y$) in both reflected and transmitted radiations, as shown in subplots (e) and (f) of Fig. \ref{fft_0p4}. It is to be noticed that, unlike a magnetized case, no harmonics with the polarization ($E_y$, $B_z$) is generated in the absence of external magnetic field, as has been shown in subplot (b) of Fig. \ref{field_0p4}. In the case with $B_0 = 0$, the existence of so-called selection rules for the polarization of harmonic generation reflected from an overdense plasma surface was predicted by \cite{lichters1996short}. Their study shows that for a normal incident linearly polarized laser, only the odd harmonics with linear polarization appear in the reflected radiation. The analogy of third-harmonic generation reported in our manuscript is the same as reported by \cite{lichters1996short}.  However, here we have observed and characterized them in both reflected and transmitted radiations. It is interesting to note that we have shown the FFT of transverse electric and magnetic field components in each subplot as a pair. This is done to show the electromagnetic nature of harmonic radiations. It is also important to observe that the polarization of third harmonic radiation is the same as the incident laser pulse. In contrast, the polarization of the second harmonic radiation is different from that of the incident laser pulse. In subsequent subsection \ref{mechansm}, we will discuss the reason behind this.
 
 The conversion efficiencies of second and third harmonics have been provided in table \ref{effi} for different values of $a_0 (=eE_l/m\omega_lc)$ and laser frequency $\omega_l$. The conversion efficiency has been calculated by taking the ratio of spatially integrated electromagnetic field (emf) energy of the harmonic to that of the incident laser pulse. We have observed that the conversion efficiency solely depends on the strength of the laser fields for a given magnetic field. Keeping $a_0$ constant when we increase $\omega_l$, field strength increases, and so do the harmonics' efficiencies. On the other hand, as we keep the value of $(a_0\omega_l)$ constant for a different combination of $a_0$ and $\omega_l$, the field strength of the incident laser pulse remains the same, and so does the conversion efficiency. This has been clearly shown in table \ref{effi}.


\begin{table*}
\centering
\caption{Conversion efficiencies of harmonics in O-mode configuration of incident laser pulse for $B_0=2.5$.}

\vspace{0.2cm}
	\begin{tabular}{|p{0.8cm}||p{1cm}||p{1cm}||p{2.0cm}||p{2.2cm}||p{2.2cm}|}
		
		\hline
		\textcolor{black}{$a_0$}& \textcolor{black}{$\omega_l$}& 	\textcolor{black}{$a_0\omega_l$}& \textcolor{black}{$\eta_{2^{nd}}(ref)(\%)$} & 	\textcolor{black}{$\eta_{2^{nd}}(trans)(\%)$} & 	\textcolor{black}{$\eta_{3^{rd}}(trans)(\%)$} \\

		\hline

		0.5&0.7&0.35&0.346&0.209&0.0342\\
		\hline
		0.5&0.6&0.30&0.264&0.173&0.024\\
		\hline
		0.5&0.5&0.25&0.193&0.128&0.016\\
		\hline
			0.5&0.4&0.20&0.130&0.085&0.0094\\ 
		
		0.4&0.5&0.20&0.125&0.0833&0.009\\
	\hline
	\end{tabular}
		\label{effi}

\end{table*}

\subsection{HG in X-mode configuration $(\protect\vv{\bm{E}}_l \perp \protect\vv{\bm{B}}_0)$}
\label{hrmnics_X-mode}
 
\begin{figure*}
\centering
   \includegraphics[height = 3.0cm,width = 13.0cm]{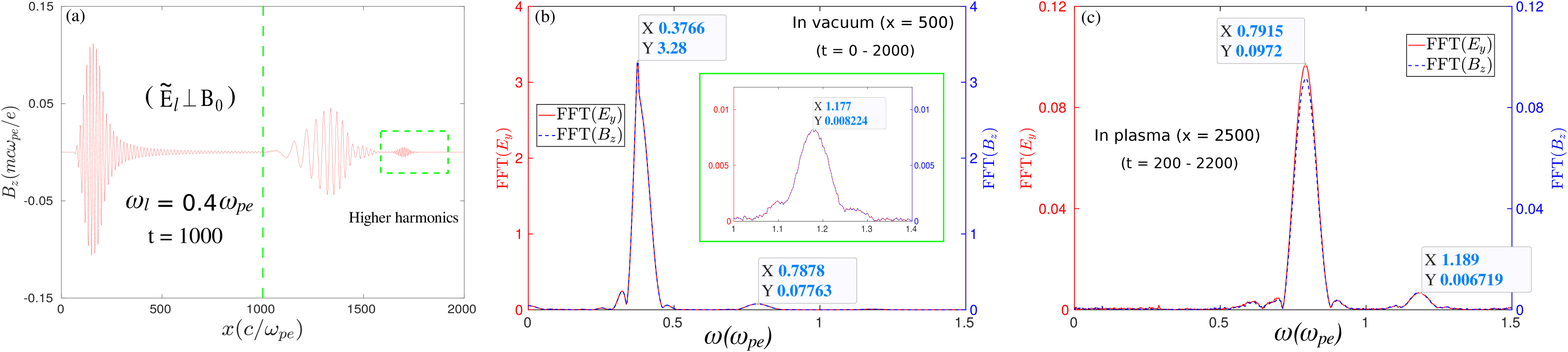}
   
   \caption{The generation of higher harmonics has been depicted here for the case where the polarization of incident laser has been chosen to be in X-mode configuration, i.e., $\mathbf{\widetilde{E}}_{l} \perp \mathbf{B}_0$. Here, we have considered $B_0 = 2.5$. The electromagnetic part of the magnetic field along $\hat z$, $B_z$ has been shown in subplot (a) at a particular instant of time $t = 1000$. In the subplot (b), the FFT of $E_y$ and $B_z$ at the location $x = 500$ (vacuum) has been sown. It is clearly seen that in addition to the original reflected laser field ($\omega \approx 0.4$), higher harmonics ($\omega \approx 0.8, 1.2$) are also present in the reflected radiation. The existence of these higher harmonics inside the bulk plasma has been depicted in subplot(c), where the FFTs have been performed at the location $x = 2500$.}

  \label{P_pol}
\end{figure*}

 The higher harmonics can also be observed for the case when the polarization of the incident laser pulse is chosen to be in X-mode configuration, i.e., $\vv{\bm{E}}_l \perp \vv{\bm{B}}_0$. This has been clearly illustrated in Fig. \ref{P_pol}. Laser pulse with frequency $0.4 \omega_{pe}$ was set up initially ($t = 0$)  to propagate in $\hat x$ direction from the location $x = 950$. In this case also, the external magnetic field $B_0$ has been chosen to be $2.5$ and applied in the $\hat z$ direction. It is to be noticed that for our chosen values of system parameters, the frequency of the incident laser pulse lies in between left-hand cutoff ($\omega_L = 0.376$) and upper hybrid frequency ($\omega_{UH} = 2.7$). Thus, the incident laser pulse with polarization in the X-mode configuration will match the passband of the plasma X-mode dispersion curve. As a result, a significant part of the incident laser pulse penetrates inside the bulk plasma. This has been clearly illustrated in subplot (a) of Fig. \ref{P_pol} where we have shown the $z-$component of the magnetic field $B_z$ as a function of  $x$ at a particular instant of time $t = 1000$. It is also seen that a part of the incident pulse gets reflected from the vacuum-plasma interface and propagates in the $-\hat x$ direction in the vacuum. The other part of the incident laser penetrates the plasma surface and propagates through the medium. It can be observed that a small disturbance, as highlighted by the dotted rectangular box in subplot (a), is also present, which moves with a higher group velocity in the plasma medium. These are essentially the higher harmonics generated by the plasma. The FFT in time for this signal of the transverse fields $E_y$ and $B_z$ in subplots (b) and (c) of Fig. \ref{P_pol} corroborate this. The FFTs of $E_y$ and $B_z$ inside the plasma ($x = 2500$), as have been shown in subplot (c), evaluated within a time window $t = 200$ to $2200$. This choice eliminates the originally transmitted laser pulse with frequency $0.4\omega_{pe}$ (having higher power) from the frequency spectrum. Two distinct peaks observed at  $\omega \approx 0.8\omega_{pe}$ and $1.2\omega_{pe}$ in both the subplots correspond to the second and third harmonic, respectively. It is interesting to notice that unlike the previous case (O-mode configuration), here, the polarization of the higher harmonics, both second and third, is the same as the incident laser pulse.

\subsection{Mechanism of HG in a magnetized plasma}
 \label{mechansm}

\begin{figure*}
\centering
   \includegraphics[height = 7.0cm,width = 12.0cm]{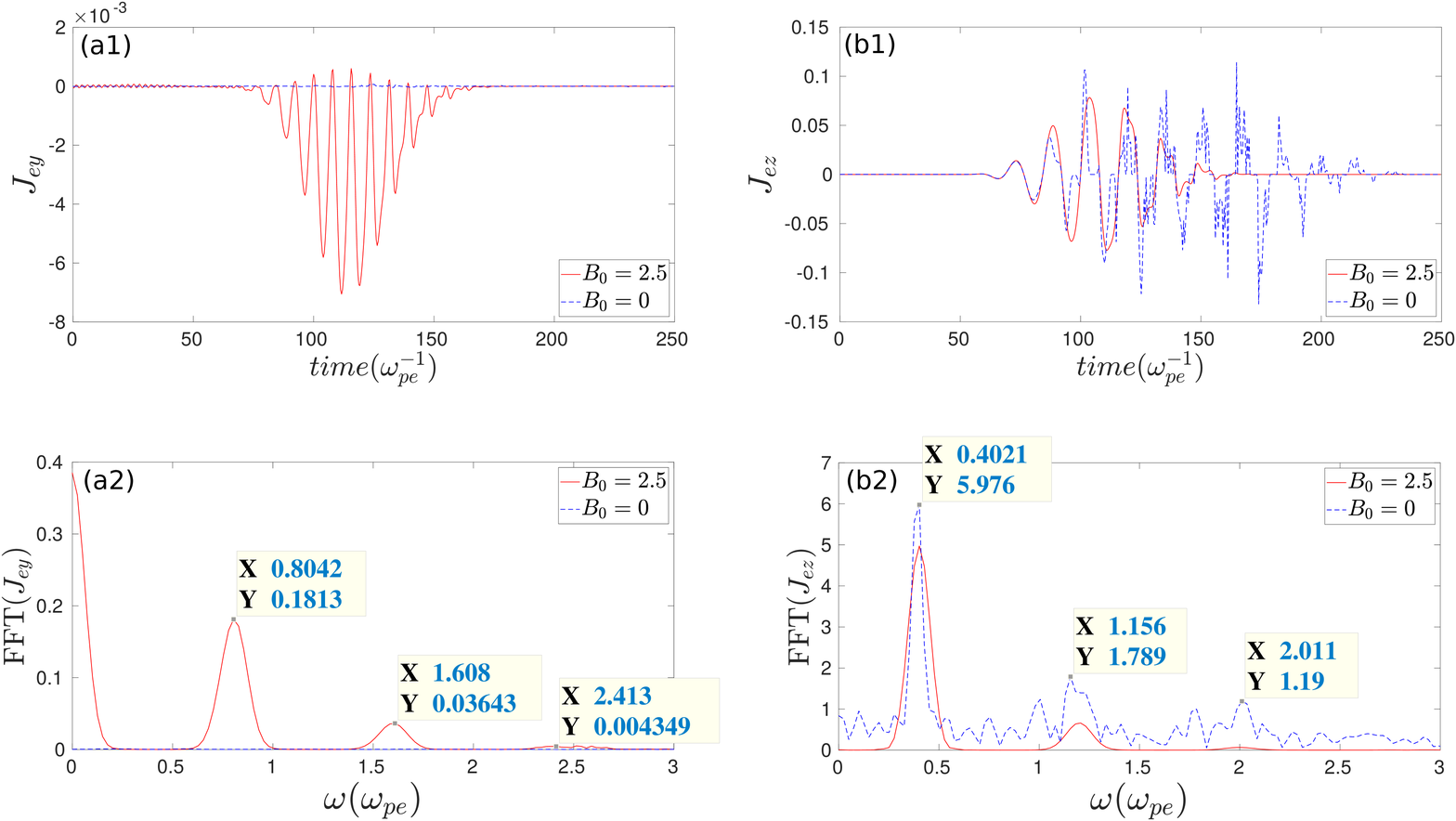}
   
   \caption{Figure shows the time evolution of (a1) $y$-component and (b1) $z$-component of electron currents $J_{ey}$ and $J_{ez}$ at the vacuum plasma interface ($x = 1000$), respectively. In subplots (a2) and (b2), the FFTs of  $J_{ey}$ and $J_{ez}$ have been shown, respectively. Here, the red solid lines and blue dotted lines represent the case with $B_0 = 2.5$ and $B_0 = 0$, respectively.}

  \label{current_fft}
\end{figure*}

 
 Let us now understand the mechanism of the harmonic generation. The fundamental mechanism of high-order harmonic generation at the vacuum-magnetized plasma interface has been previously studied by \cite{mu2016effect}. In the present work, we have briefly discussed the same extending it for both O and X-mode configurations of the incident laser pulse. Additionally, we have also developed a simple approximate mathematical model for a qualitative description of the observed characteristics in both O and X-mode configuration, and provided in the appendix. 
 
 When a laser pulse with  O-mode configuration, i.e., $\vv{\bm{E}}_l \parallel \vv{\bm{B}}_0$ is  incident on the vacuum-plasma interface, plasma particles will experience a force ($\propto \widetilde{v}_{z}\widetilde{B}_{l}\exp({i2\omega_l t)}$) due to the Lorentz force ($\vv{\bm{v}} \times \vv{\bm{B}}_l$) along $\hat x$ direction. Here, $\vv{\bm{v}}$ is the particle's quiver velocity initiated due to the laser electric field $\bm{E_l}$ in $\hat z$. As a result, plasma electrons wiggle, forming an oscillating current at the surface of the plasma in $\pm \hat x$ direction with a frequency twice the incident laser frequency (Eq. \ref{current_xy} ). This can also be understood by recalling that in the intense EM fields of incident laser, electron motion would initially follow the well known ``figure-of-8'' path, which contains a transverse component with a frequency $\omega_l$ and a longitudinal part with $2\omega_l$ frequency \cite{Rev_harmonic}. The electrons being magnetized in the presence of the external magnetic field $B_0\hat{z}$, revolve in the $x-y$ plane. Thus, the oscillatory motion of electrons along $\hat x$ generated by the process discussed above is coupled with its motion along $\hat{y}$,  having the same frequency. Consequently, an oscillating current with a frequency twice the laser frequency is produced in the $\hat y$ direction, as have been shown in subplots (a1) and (a2) of Fig. \ref{current_fft}. This acts as an oscillating current sheet antenna and radiates electromagnetic wave with the fields $\widetilde{E}_y$ and $\widetilde{B}_z$ propagating along both $\pm \hat x$ directions. These are the second harmonics that have been captured in our simulations in both vacuum and plasma regimes. It is easy to understand that the oscillating current sheets in the $\hat y$ direction might have all the even higher harmonics depending upon the strength of the non-linearity involved in the electron dynamics. In our simulations, we have captured up to sixth harmonic and can be easily seen in the FFT $J_{ey}$ in subplots (a2) of Fig. \ref{current_fft}. 
  
  In addition,  the electron motion along $\hat x$   will couple to the laser magnetic field $\widetilde{B}_l$ (which is along $\hat y$ for O-mode configuration). As a result, it will create an oscillating current sheet along $\hat z$ with the frequency thrice of laser frequency and also at higher odd harmonic values, as have been shown in subplots (b1) and (b2) of Fig. \ref{current_fft}. Thus, another radiation will be produced propagating in $\pm x$ direction, but in this case, the electromagnetic fields associated with this radiation are $\widetilde{E}_z$ and $\widetilde{B}_y$. It is to be noticed that ideally, all the odd higher harmonics might be present in the $J_{ez}$. In our simulations, we have identified up to fifth harmonic and shown in subplot (b2) of Fig .\ref{current_fft}. It is to be noticed that the external magnetic field is necessary to generate the $\hat y$-component of surface current $J_{ey}$, oscillating with the even harmonic frequencies, producing the even higher harmonic EM radiations. This has been shown by red solid and blue dotted lines in the subplot (a1) and (a2) of Fig. \ref{current_fft} and is also consistent with the approximate mathematical expression of $J_{ey}$ in Eq. \ref{current_xy}. On the other hand, the $\hat z$ component of surface current, \emph{i.e.} $J_{ez}$ oscillating with odd harmonic frequencies is produced only due to the nonlinear electron dynamics in the laser fields and does not require any external magnetic field \cite{lichters1996short, Rev_harmonic, bulanov1994interaction}, as has been shown in the subplots (b1) and (b2) of Fig. \ref{current_fft}. This is also apparent from the expression of $J_{ez}$ in Eq. \ref{jz3}.  
 
 For the laser polarization in the X-mode configuration,  the magnetic field $\widetilde{B}_l$ of the laser is parallel to ${B_0}$ (along $\hat z$). For this case, using similar arguments, both second and third harmonics (in fact, all the high harmonics) will be generated for which the electromagnetic fields are $\widetilde{E}_y$ and $\widetilde{B}_z$. Thus, the generated harmonic radiation also has the X-mode configuration. This is exactly what we have observed in our simulations, as have been shown in Fig. \ref{P_pol}.

\begin{figure*}
\centering
 \includegraphics[height = 3.0cm,width = 13.0cm]{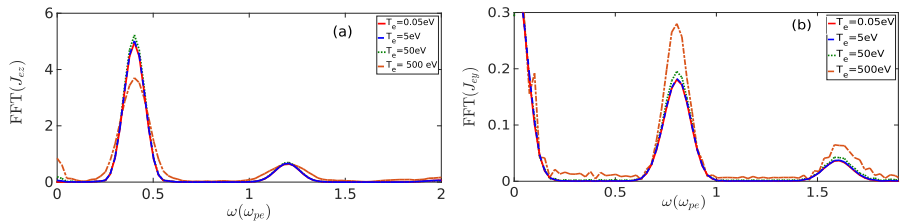}
   
   \caption{FFT in time for (a) $\hat{z}$-component of surface current $J_{ez}$ and (b) $\hat{y}$-component of surface current $J_{ey}$ at the vacuum-plasma interface ($x=1000 c/\omega_{pe}$) for different electron temperature ($T_e=0.05eV, 5eV, 50eV$ and $500eV$).}

  \label{temp_hg}
\end{figure*}

 As mentioned in section \ref{picsim}, in most of our simulation studies we have considered the electron temperature to be very low ($T_e = 0.05eV$). However, we have also done a comparative study by considering different values of electron temperature to check if the fundamental mechanism of HG has any dependence on the plasma temperature. The time FFT of surface currents $J_{ez}$ and $J_{ey}$ has been shown for different values of electron temperature ($T_e = 0.05eV-500eV$) in subplots (a) and (b) of Fig. \ref{temp_hg}, respectively. It is seen that surface currents oscillating with high harmonic frequencies have been generated in all the cases. There is no significant change observed in the generation of high harmonic radiations for different values of electron temperature. 
 
 \subsection {Effect of external magnetic field on HG}
 \label{varying_B0}

\begin{figure*}
\centering
\includegraphics[height = 7.5cm,width = 9.5cm]{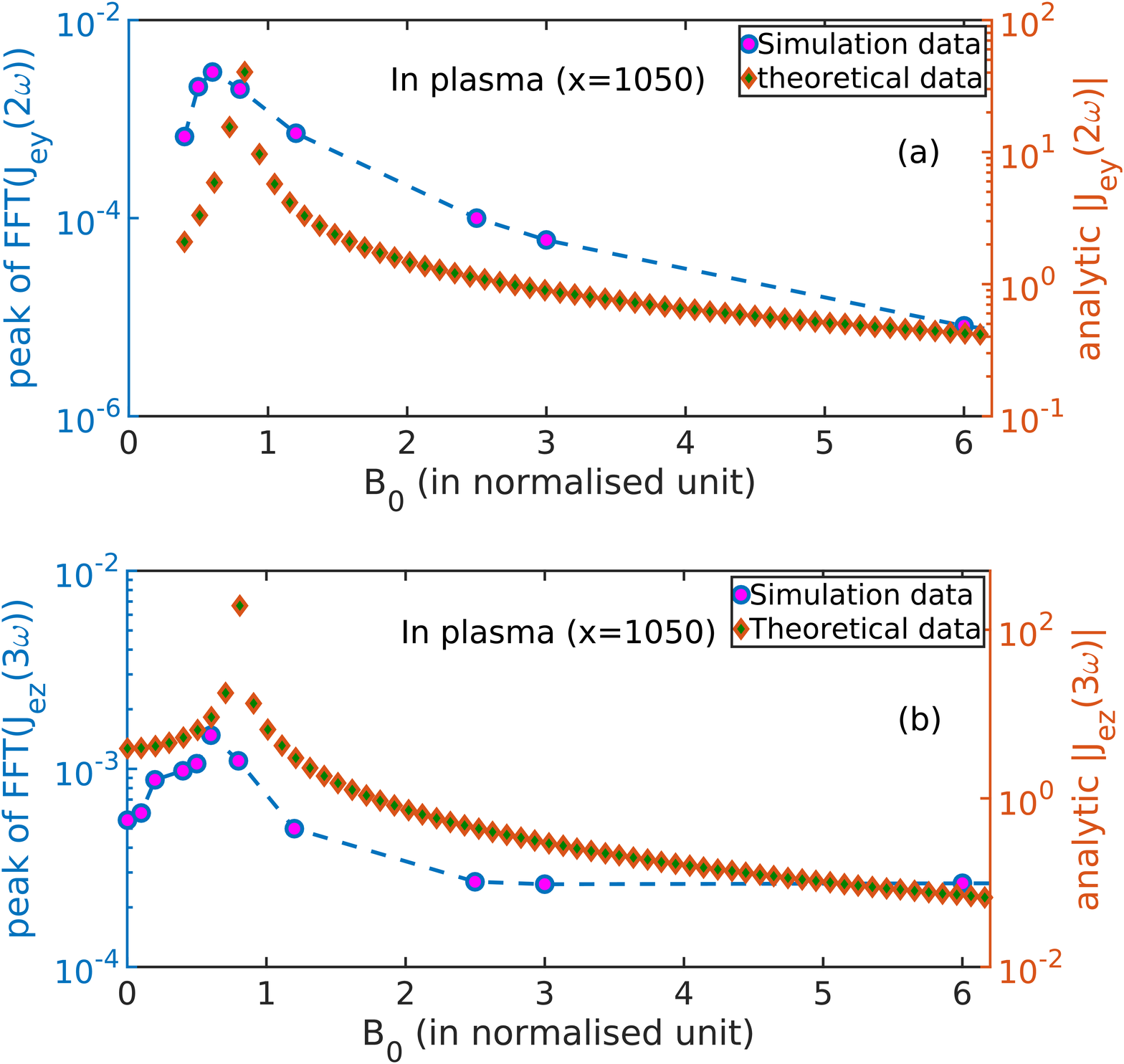}
   \caption{(a) The peak value of the FFT spectrum of $J_{ey}$ corresponding to the second harmonic frequency and theoretical value of $|{J_{ey}}|$ obtained from the Eq. \ref{current_xy}. (b) The peak value of the FFT spectrum of $J_{ez}$ corresponding to the third harmonic frequency and theoretical value of $|{J_{ez}}|$ obtained from the Eq. \ref{jz3}.}

  \label{compjB}
\end{figure*}

We have shown in the previous section that the generation of harmonics depends on the surface current oscillations, and odd harmonics can be generated even when there is no external magnetic field. However, it has been shown that an external magnetic field is necessary to generate higher even harmonics. In this section, we will show that the amplitude of harmonics (both even and odd) actually has a strong dependence on applied external magnetic fields.

For this analysis, we have done a comparative study by simulating the same geometry with $\omega_l=0.4\omega_{pe}$ for different values of the external magnetic field. In each run, we have obtained the amplitude (peak value) of FFT from time-series data of $J_{ey}$ and $J_{ez}$ corresponding to the second and third harmonic frequency, respectively. Next, we have calculated the absolute value of second and third harmonic current density from the approximated model given in appendix \emph{i.e.} Eq. \ref{current_xy} and \ref{jz3} for different magnetic field values. We have plotted these two quantities as a function of the external magnetic field $B_0$ and shown in Fig. \ref{compjB}. One can observe that trend is similar for quantities obtained from simulation and theoretical model. This plot provides a qualitative understanding of the mechanism presented in this paper. This observation also establishes that harmonic generation in plasma is boosted in the presence of an external magnetic field. There is an optimum value of the magnetic field for which better efficiency of harmonic generation can be found, and this value is where $\omega_{ce} \rightarrow 2\omega_{l}$.

\subsection{Characterization of harmonics}
\label{chh}

\begin{figure*}
\centering
   \includegraphics[height = 5.5cm,width = 13.0cm]{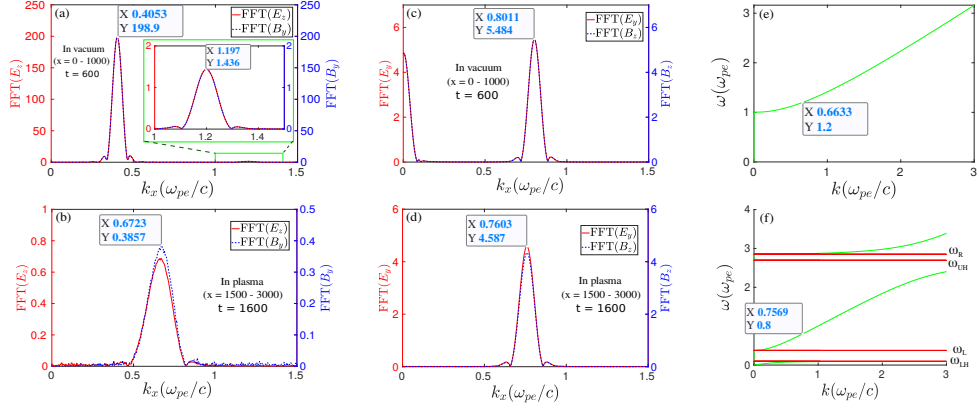}
   
   \caption{Fourier transforms in space of the electromagnetic fields after the laser is reflected from the vacuum-plasma interface. In subplots (a) and (b), the FFT of ($E_z$, $B_y$) along $\hat x$ at time $t = 600$ and $1600$ have been shown for vacuum and bulk plasma, respectively. On the other hand, the same has been shown for the fields ($E_y$, $B_z$) in subplots (c) and (d), respectively. Dispersion curves (\cite{boyd2003physics}) of ordinary ($O$) mode and extraordinary ($X$) mode for the chosen values of the system parameters of this study have been shown in subplots (e) and (f), respectively.}

  \label{space_fft}
\end{figure*}

\begin{figure*}
\centering
   \includegraphics[height = 3.0cm,width = 13.0cm]{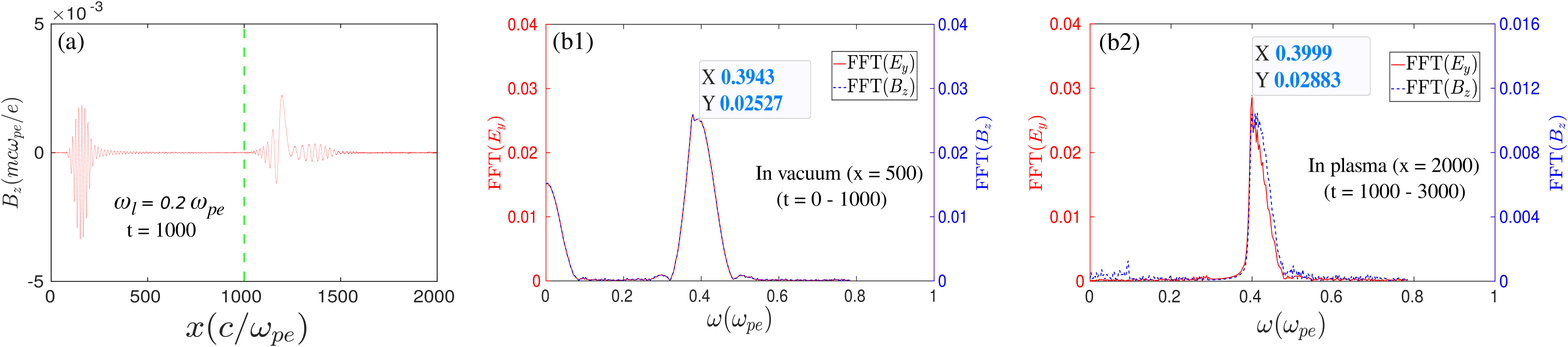}
   
   \caption{The z-component of the magnetic field $B_z$ (electromagnetic) with respect to $x$ at time $t = 1000$ has been shown in subplot (a). The green dotted line at the location $x = 1000$ represents the vacuum-plasma interface. The FFTs of ($E_y$, $B_z$) at the locations $x = 500$ and $x = 2000$ in the frequency domain, as have been shown in subplots (b1) and (b2), clearly demonstrate that second harmonic is present in both reflected and transmitted radiation, respectively.}

  \label{freq_modeX}
\end{figure*}

\begin{figure*}
\centering
   \includegraphics[height = 4.0cm,width = 13.0cm]{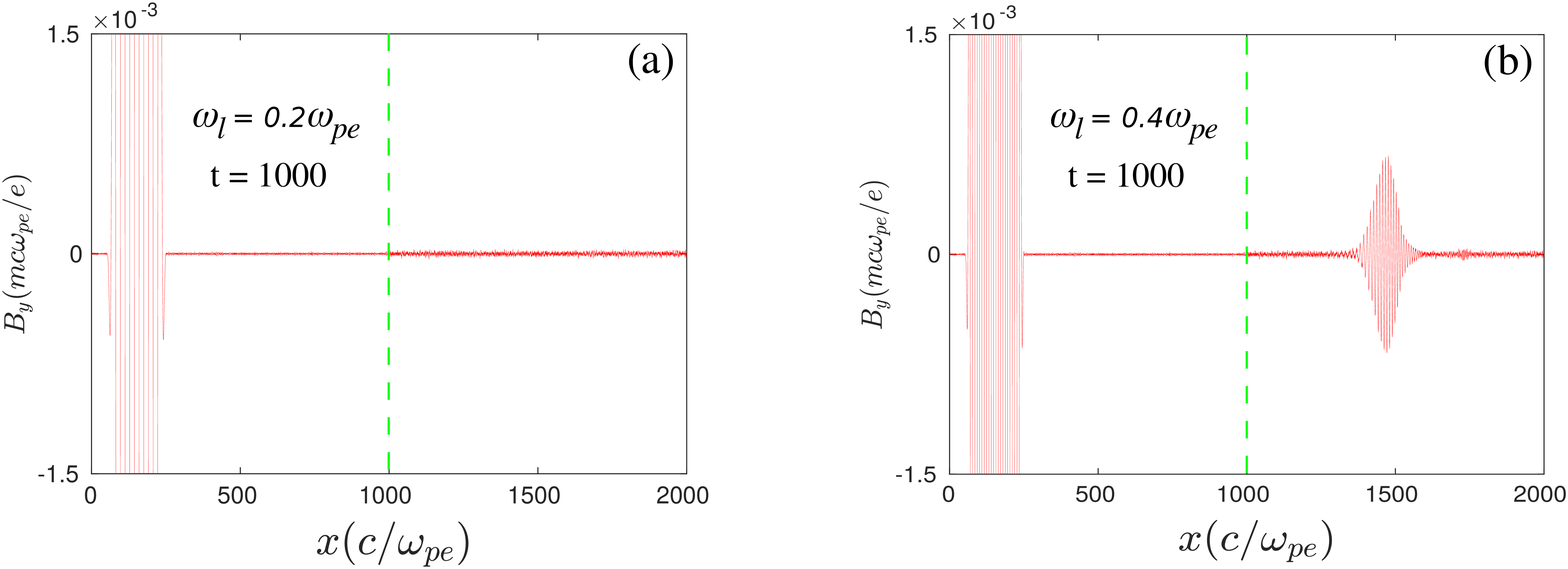}
   
   \caption{ The y-component of magnetic field $B_y$ (electromagnetic) with respect to $x$ at a particular instant of time $t = 1000$ has been shown for incident laser frequency (a) $\omega_l = 0.2$ and (b) $\omega_l = 0.4$. In both the cases, the polarization of the incident laser has been chosen to be in O-mode configuration, i.e., $\mathbf{\widetilde{E}}_{l} \parallel \mathbf{B}_0$ (along $\hat z$) and the value of $a_0$ is considered to be 0.5. Here, the green dotted line at $x = 1000$ represents the vacuum-plasma interface.}

  \label{freq_modeO}
\end{figure*}

\begin{figure*}
\centering
   \includegraphics[height = 3.0cm,width = 14.0cm]{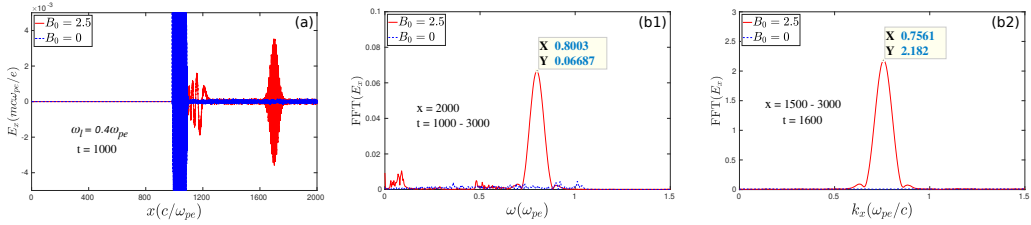}
   
   \caption{The longitudinal electric field $E_x$ with respect to $x$ has been shown in the subplot (a) at a particular instant of time $t = 1000$. In the subplots (b1) and (b2), the FFTs of $E_x$ in frequency and $k-$space have been shown, respectively. Here, the red solid lines and blue dotted lines represent the case with $B_0 = 2.5$ and $B_0 = 0$, respectively.}

  \label{Efield_x}
\end{figure*}


\begin{figure*}
\centering
   \includegraphics[height = 4.0cm,width = 12.0cm]{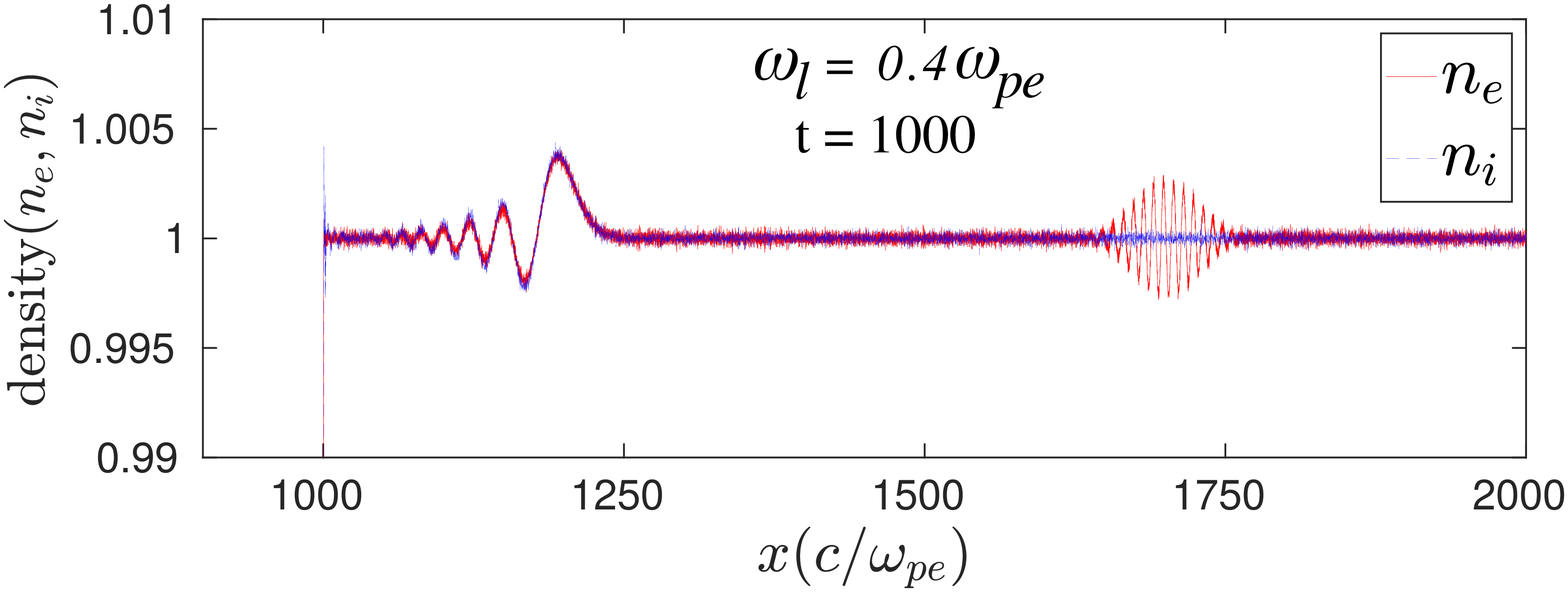}
   \caption{The electrons and ions density fluctuations have been shown by red solid and blue dotted lines, respectively. Here, the external magnetic field $B_0$ is considered to be $2.5$.}

  \label{densty}
\end{figure*}

We now analyze in further detail and characterize the high harmonic radiations.  As has been shown previously, these higher harmonics can be observed for the laser polarization in both O-mode ($\vv{\bm{E}}_l \parallel \vv{\bm{B}}_0$) and X-mode ($\vv {\bm{E}}_l \perp \vv{\bm{B}}_0$) configurations. We consider here the observations corresponding to the O-mode configuration. As discussed in the previous sections, the higher harmonic radiations are electromagnetic. Thus, in the vacuum, these radiations will travel with the speed of light, i.e., the frequency and wavenumber (in normalized units) will have the same values. On the other hand, when they propagate through the plasma medium, they will have to follow the appropriate dispersion relation of the plasma medium. 

Thus, the group velocity and the phase velocity of these radiations will have different values depending upon the dispersion relation. These properties have been clearly shown in Fig. \ref{space_fft}. The spatial FFTs of the transverse fields ($E_z$, $B_y$) for vacuum and bulk plasma have been shown in subplots (a) and (b) of Fig. \ref{space_fft}, respectively. It is clearly seen from subplot (a) that, as expected, the reflected laser pulse ($\omega_l = 0.4$) and the third harmonic ($\omega \approx 1.2$) radiation are associated with the wavenumbers $k_x \approx 0.4$ and $k_x \approx 1.2$, respectively, as they travel in vacuum. On the other hand, the spectrum of the spatial FFTs of $E_z$ and $B_y$ in the bulk plasma region shows a distinct peak at a particular value of $k_x \approx 0.67$. It is interesting to realize that the third harmonic radiation is associated with the transverse electric field ($E_z$) parallel to the external magnetic field $B_0$ and traveling perpendicular to $B_0$. Thus, it matches the condition of plasma ordinary ($O$) mode. We have evaluated the theoretical dispersion curve for the $O$ mode for our chosen values of system parameters and have shown in the subplot (e) of Fig. \ref{space_fft}. It is seen that the value of wavenumber corresponding to the frequency $\omega = 1.2$ is approximately equal to $0.66$. Thus, It matches well with the properties of third harmonic radiation observed in our simulation inside the bulk plasma region.

 The FFT of transverse fields $E_y$ and $B_z$ in space associated with the second harmonic radiation have been depicted in subplots (c) and (d) of Fig. \ref{space_fft} for vacuum and bulk plasma, respectively. The FFT spectrum reveals that the second harmonic radiation ($\omega \approx 0.8$) propagates with a finite wavenumber in a vacuum, $k_x \approx 0.8$, which is expected as it has to travel with the speed of light. On the other hand, inside the plasma, it is associated with a different wavenumber $k_x \approx 0.76$. The theoretical model analysis again affirms that this second harmonic radiation matches the condition for plasma extraordinary ($X$) mode and propagates through the passband lying in between left-hand cutoff ($\omega_L$) and upper hybrid frequency ($\omega_{UH}$) of the $X$ mode dispersion curve. This has been demonstrated in the subplot (f) of Fig. \ref{space_fft}.

The mode analysis of harmonic radiations has a direct significance in the sense that we can now have control over the excitation of these radiations inside the plasma medium. For a given set of plasma parameters, if we change the value of $B_0$, the dispersion curves of the plasma modes (Fig. \ref{space_fft}) will be modified accordingly. Thus, the value of $B_0$ determines whether harmonic radiation with a particular frequency generated at the vacuum-plasma interface will be able to transmit inside the plasma or not. Alternatively, if the value of $\omega_l$  is changed, the frequency of the higher harmonics will also be modified accordingly. Thus, there might be some situations where these harmonics will not be allowed to pass through the plasma medium for a particular value of $B_0$. For instance, we have considered a particular case where the frequency of the incident laser pulse is chosen to be $0.2 \omega_{pe}$ and all other system parameters have been kept the same as the previous case. It has been observed that the second harmonic radiations initiated at the vacuum-plasma interface are traveling in both vacuum and plasma. This has been clearly demonstrated in the subplot (a) of Fig. \ref{freq_modeX}. It is to be noticed that the frequency of the second harmonic radiation, which happens to be $0.4\omega_{pe}$ in this case (subplots (b1) $\&$ (b2)), is still higher than the left-hand cutoff $\omega_L = 0.376$ and thus, lies within the passband region in between $\omega_L$ and $\omega_{UH}$. Hence, the second harmonic radiation is allowed to pass through the plasma. On the other hand, the third harmonic radiation, which has the O-mode characteristics, will have the frequency $0.6\omega_{pe}$. Thus, in this particular case, the third harmonic radiation is lying below the cutoff ($\omega = \omega_{pe}$) of the O-mode dispersion curve and is forbidden to propagate inside the plasma. This has been clearly illustrated in Fig. \ref{freq_modeO}. In subplot (a) and (b) of Fig. \ref{freq_modeO}, we have shown the y-component of the transverse magnetic field $\widetilde{B}_y$  at a particular instant of time $t = 1000$ for two different incident laser frequencies $\omega_l = 0.2\omega_{pe}$ and $0.4\omega_{pe}$, respectively. It is seen that for $\omega_l = 0.2\omega_{pe}$ (subplot (a)), electromagnetic field $\widetilde{B}_y$ of the incident laser pulse has been reflected from the vacuum-plasma interface and no signal of $\widetilde{B}_y$ exists inside the plasma. Whereas, for $\omega_l = 0.4\omega_{pe}$ (subplot (b)), a part of $\widetilde{B}_y$ is also present inside the plasma and which is associated with the third harmonic radiation, as also has been demonstrated in the section \ref{hrmnics_Omode}.


It is straightforward that in the presence of an external magnetic field $B_0$, a plasma wave with the electric field $\vv{\bm{E}} \perp \vv{\bm{B}}_0$ and the propagation vector $\vv{\bm{k}} \perp \vv{\bm{B}}_0$ (X-mode configuration) always tend to be elliptically polarized instead of plane-polarized \cite{chen1984introduction}. That is, when an EM wave propagates through the plasma, an electric field component parallel to the propagation direction will also be present. The wave, therefore,  has both electromagnetic and electrostatic features. In our study, the observed second-harmonic radiation is associated with an electric field perpendicular to $B_0$ and propagates along $\hat x$, as shown in previous sections. Thus, an electric field component along $\hat x$, $\widetilde{E}_x$ is also expected to be present and will be traveling along with the second harmonic EM radiation. This has been clearly depicted in the subplot (a) of Fig. \ref{Efield_x}. In this subplot, the $E_x$ field profile is seen to be associated with the second harmonic radiation. The Fourier spectra of this particular profile in both $\omega$ and $k$-space also show the same characteristic properties as the electromagnetic second harmonic radiation, as has been shown in subplots (b1) and (b2) of Fig. \ref{Efield_x}. It is to be noticed that no $\hat x$-component of electric field ($E_x$) is observed inside the bulk plasma for the case without external magnetic field, as has been shown by blue dotted lines in Fig. \ref{Efield_x}. This is consistent with the results discussed in section \ref{hrmnics_Omode} and section \ref{mechansm}. In this case, $E_x$ is present only at the vacuum-plasma interface where the disturbances were made due to the ponderomotive pressure of the incident laser pulse. It can not propagate inside the bulk plasma without an external magnetic field. It should be noted that the odd harmonics getting generated and present inside the bulk plasma for $B_0 = 0$ (Fig. \ref{field_0p4}) can also induce longitudinal electric field $E_x$. However, this will be negligible for $B_0 = 0$ as the longitudinal electric field will, in this case, get generated by the coupling of dynamics of electrons in the electric and magnetic field of the harmonic EM wave radiation ( i.e., $\vec{E}_{nh}\times\vec{B}_{nh}$, where, $n$ represents the order of harmonic). Since the intensity of the harmonic radiation is typically much weaker, this effect will be very small compared to  $E_x$ that gets generated by the coupling of the electric field of the harmonic radiation with the external magnetic field.   
This has been clearly illustrated in Fig. \ref{Efield_x}.

 We also observe from the subplot (a) of Fig. \ref{Efield_x} that a large scale disturbance (red solid line) is present in $E_x$ near the plasma-vacuum interface for $B_0 = 2.5$. Such a disturbance has also been observed in the transverse fields $E_y$ and $B_z$, as has been shown in subplot (b) of Fig. \ref{field_0p4}. This fluctuation is electromagnetic but becomes elliptically polarized in the presence of an external magnetic field $B_0$, as discussed earlier. It is also observed that this disturbance travels with a much slower velocity than the higher harmonic radiation present in the system. Such a disturbance has also been observed in both electron and ion density profiles, as can be seen in Fig. \ref{densty}. The normalized density profiles of electrons and ions at a particular instant of time $t = 1000$ have been shown by the solid red line and blue dotted line in Fig. \ref{densty}, respectively. A spatial electron density profile travels along with the generated second harmonic pulsed structure wave and is thus associated with it. As the second harmonic frequency is much higher than the ion response time scale, this structure appears only in the electron density profile, not in ions. On the other hand, near the plasma surface, the fluctuations are present in both electron and ion density profiles, and they are almost identical. These fluctuations are associated with low-frequency magnetosonic perturbations. They were initiated due to the ponderomotive force associated with the finite pulse width of the laser. However, for the choice of infinitely massive ions, these magnetosonic perturbations do not appear. Such a disturbance has also been reported in a recent study by \cite{ayushi_nf} for  X-mode configuration. We confirm that this is also present in the O-mode configuration.


\section{Summary}
\label{smry}
We have shown that the interaction of laser with a magnetized plasma has interesting and rich facets. 
The O-mode geometry, which is typically considered to exhibit unmagnetized plasma response, has certain distinctive features. Specifically, we have 
demonstrated the generation of  harmonics in a 
  magnetized plasma even in the O-mode configuration for overdense plasma regime. The dynamics of a laser pulse interacting with plasma have been followed up by one-dimensional PIC simulations using OSIRIS-4.0. A laser pulse coming from a vacuum is chosen to fall on an overdense plasma medium ($\omega_l<\omega_{pe}$) in the presence of an externally applied magnetic field. The dynamical  mechanisms leading to higher harmonic generation in both O-mode  ($\vv{\bm{E}}_l \parallel \vv{\bm{B}}_0$) and X-mode configurations ($\vv{\bm{E}}_l \perp \vv{\bm{B}}_0$) have been demonstrated and analyzed. It has been shown that when the incident laser has the polarization in the O-mode configuration, the higher even harmonics will be produced in the X-mode configuration, and odd harmonics will be in the O-mode configuration. Whereas both even and odd harmonics will be in the X-mode configuration for the incident laser pulse in the X-mode configuration. A comparison of simulation results with and without an external magnetic field has been provided and discussed. The required conditions for the propagation of the harmonic radiation inside the plasma have been identified. Our study reveals that the conversion efficiency for the harmonic generation increases with laser intensity. We have also shown that the conversion efficiency has a strong dependence on the externally applied magnetic field. There is an optimum value of the magnetic field for which the amplitude of harmonics is maximum. Our study also demonstrates a conversion efficiency of about  $1.77\%$ for second harmonic for a laser intensity of $a_0=0.5$ and $B_0=0.6$, which is higher than the previously reported value in the analytical studies done in the context of magnetized plasma (\cite{jha2007second, ghorbanalilu2012second}). We feel that this can be improved further by appropriate tailoring of the plasma and magnetic field profiles.

\section{Acknowledgements} 
The authors would like to acknowledge the OSIRIS Consortium, consisting of UCLA and IST (Lisbon, Portugal) for providing access to the OSIRIS 4.0 framework which is the work supported by NSF ACI-1339893. This research work has been supported by the 
 J. C. Bose fellowship grant of AD (JCB/2017/000055) and
the CRG/2018/000624 grant of DST. The authors thank IIT Delhi HPC facility for computational resources.

\onecolumngrid
\appendix

\section{Surface current oscillation in magnetized configuration}
\noindent
We demonstrate the possible generation of harmonics by the surface electron currents, which would get excited by the laser field falling on an overdense plasma target in both O and X - mode configuration. The spatial variation along $x$ is ignored for simplicity. \\
\subsection{Surface current oscillation in O-mode configuration}
In the O-mode configuration of incident laser wave, we have the external magnetic fields and the electromagnetic (EM) fields of an incident wave as follows: 

\begin{eqnarray}
\vv{B_0}=B_0\hat{z}\hspace{0.5cm};\hspace{0.5cm} 
\vv{\widetilde{E}_l}=E_{lz}exp(-i \omega_{l}t)\hat{z} \hspace{0.5cm};\hspace{0.5cm}
\vv{\widetilde{B}_l}=\frac{E_{lz}}{c} exp(-i \omega_{l}t)\hat{y}
\end{eqnarray}
The electron velocity is expanded in terms of various orders  in  the laser field amplitude as 

$$\vv{\widetilde{v}} = \vv{\widetilde{v}}^{(1)} + \vv{\widetilde{v}}^{(2)} + \vv{\widetilde{v}}^{(3)} + ...$$

\noindent
From the Lorentz force equation we obtain for electrons 
 
\begin{equation}
\frac{\partial{\vv{\widetilde{v}}}}{\partial t}=-\frac{e}{m} \left[\vv{\widetilde{E}_l}+\vv{\widetilde{v}}\times(\vv{B_0}+\vv{\widetilde{B}_l})\right]
\label{Lorntz}
\end{equation}

Balancing the  first order linear term in  Eq. \ref{Lorntz} we obtain  for the $\hat{z}$ component 

\begin{eqnarray*}
\frac{\partial \widetilde{v}_{z}^{(1)}}{\partial t}=-\frac{e}{m}\widetilde{E}_l \hspace{0.5cm};\hspace{0.5cm}
\widetilde{v}_{z}^{(1)}=-\frac{e}{m}\frac{i}{\omega_{l}}\widetilde{E}_l
\end{eqnarray*}

\begin{equation}
\widetilde{v}_{z}^{(1)}=-\frac{i e}{m\omega_{l}}E_{lz}exp(-i \omega_{l}t)
\end{equation}
The x and y components represents the gyromotion and depends on  the thermal velocity  of electrons and are independent of laser field. 
%
For a cold plasma this can be ignored compared to the quiver velocity of the electron in the laser field. 
 Thus, for the O-mode configuration ($\vv{\widetilde{E_1}} \perp \vv{B_0}$), $\widetilde{v}_{x}^{(1)}$ and $\widetilde{v}_{y}^{(1)}$ can be neglected. \\

\noindent
The  second order term from Eq. \ref{Lorntz} 
\begin{equation}
\frac{\partial \widetilde{v}_{z}^{(2)}}{\partial t}=-\frac{e}{m}(\widetilde{v}_{x}^{(1)}\widetilde{B}_l)\approx 0 \hspace{0.5cm}
\end{equation}

\begin{equation}
\frac{\partial \widetilde{v}_{x}^{(2)}}{\partial t}=\frac{e}{m}\left[\widetilde{v}_{z}^{(1)}\widetilde{B}_l-\widetilde{v}_{y}^{(2)}B_0\right]
\hspace{0.5cm};\hspace{0.5cm}
\frac{\partial \widetilde{v}_{y}^{(2)}}{\partial t}=\frac{e}{m}(\widetilde{v}_{x}^{(2)}B_{0})
\label{xy2}
\end{equation}

From the coupled expressions in Eq. \ref{xy2}, we will get 

\begin{eqnarray}
\frac{\partial^2 \widetilde{v}_{x}^{(2)}}{\partial t^{2}}+\omega_c^2\widetilde{v}_{x}^{(2)}=\frac{e}{m}\left[\frac{\partial(\widetilde{v}_{z}^{(1)}\widetilde{B}_l)}{\partial t}\right]
\hspace{0.5cm};\hspace{0.5cm}
\frac{\partial^2 \widetilde{v}_{y}^{(2)}}{\partial t^{2}}+\omega_c^2\widetilde{v}_{y}^{(2)}=\frac{e}{m}\omega_c\frac{(-i e)}{(m\omega_{l})}\widetilde{E}_l\widetilde{B}_l
\label{x2y2}
\end{eqnarray}

Here, $\omega_c = \frac{eB_0}{m}$. Separating the time dependence we have 
 
\begin{eqnarray*}
\widetilde{v}_{x}^{(2)}\approx v_x^{(2)}exp(-i 2\omega_{l}t) 
\hspace{0.5cm};\hspace{0.5cm}
\widetilde{v}_{y}^{(2)}\approx v_y^{(2)}exp(-i 2\omega_{l}t)
\end{eqnarray*}

\noindent
Solving the Eq. \ref{x2y2},  obtain the  expressions of the  $x$ and $y$ component of current as: 

\begin{equation}
\widetilde{J}_{ey}^{(2)}=in_ee \left(\frac{e}{m}\right)^2\frac{\omega_c}{\omega_{l}}\frac{E_{lz}^{2}}{c\left(\omega_c^2-4\omega_{l}^2\right)}exp(-i2\omega_{l}t)
\hspace{0.2cm} ; \hspace{0.2cm}
\widetilde{J}_{ex}^{(2)}=2n_ee\left(\frac{e}{m}\right)^2\frac{E_{lz}^2}{c\left(\omega_c^2-4\omega_{l}^2\right)}exp(-i 2\omega_{l}t)
\label{current_xy}
\end{equation}

\noindent
Eq. \ref{current_xy} represent the expression of currents in the $x-y$ plane oscillating with a frequency twice of incident wave (second harmonic). 
This oscillating current acts as an antenna and radiates EM radiation at the second harmonic frequency. \\

\noindent
Similarly, various components of the third order velocity  terms in  Eq. \ref{Lorntz} obey \\
\begin{equation}
\frac{\partial \widetilde{v}_{x}^{(3)}}{\partial t}=-\omega_{c}\widetilde{v}_{y}^{(3)}
\hspace{0.5cm};\hspace{0.5cm}
\frac{\partial \widetilde{v}_{y}^{(3)}}{\partial t}=\omega_{c}\widetilde{v}_{x}^{(3)}
\label{gyration_3rd}
\end{equation}

\noindent
Eq. \ref{gyration_3rd} represents  the cyclotron motion in the plane ($x-y$) perpendicular to $B_0$ with an amplitude associated with the third order correction of perpendicular velocity which can only have thermal contribution. Thus, as mentioned previously, we can neglect this for our case. \\

The $z$ component of Eq. \ref{Lorntz} for third order term will give
\begin{equation}
\frac{\partial{\widetilde{v}_{z}^{(3)}}}{\partial t}=-\frac{e}{m}\left[\widetilde{v}_{x}^{(2)}\widetilde{B}_l\right]
\end{equation} 

Using the expression of $\widetilde{v}_{x}^{(2)}$ from Eq. \ref{current_xy} it can be shown that,

\begin{equation}
\widetilde{J}_{ez}^{(3)}=-i2n_{e}e \left(\frac{e}{m}\right)^3\frac{E_{lz}^{3}}{3\omega_{l}c^2\left(\omega_c^2-4\omega_{l}^2\right)}exp(-i3\omega_{l}t)
\label{jz3}
\end{equation}

\noindent
Eq. \ref{jz3} represents the $z$ component of current  $\widetilde{J}_z$ oscillating with $3\omega_{l}$ frequency(third harmonic).\\ 

\noindent 
It is to be noted that for O-mode configuration of the incident wave the second harmonic radiation has the polarization in x-y plane ($\widetilde{J}_x$ and $\widetilde{J}_y$) and the polarization of third harmonic radiation is  along $\hat z$ $(\widetilde{J}_z$). If we calculate for all the higher order terms, it can be shown that for O-mode configuration, all the even harmonics will have the polarization in the x-y plane and all the odd harmonics will have polarization along $\hat z$. This has also been observed in our simulation, as has been shown in Fig. 5.


\subsection{Surface current oscillation in X-mode configuration}

In the X-mode configuration of incident laser wave, we have the directions of external magnetic fields and the EM fields of incident wave as follows:

\begin{eqnarray}
\vv{B_0}=B_0\hat{z}\hspace{0.5cm};\hspace{0.5cm} 
\vv{\widetilde{E}_l}=E_{ly}exp(-i \omega_{l}t)\hat{y} \hspace{0.5cm};\hspace{0.5cm}
\vv{\widetilde{B}_l}=\frac{E_{ly}}{c} exp(-i \omega_{l}t)\hat{z}
\label{x-mode}
\end{eqnarray}

\noindent
We will calculate the expression of higher order terms of currents using Eq. \ref{Lorntz} and \ref{x-mode}.\\

Solving for the first order term of Eq. \ref{Lorntz},

\begin{eqnarray}
\frac{\partial \widetilde{v}_{z}^{(1)}}{\partial t}=0\hspace{0.5cm};\hspace{0.5cm}
\frac{\partial \widetilde{v}_{y}^{(1)}}{\partial t}=-\frac{e}{m}\left[\widetilde{E}_l-\widetilde{v}_{x}^{(1)}B_0\right]\hspace{0.5cm};\hspace{0.5cm}
\widetilde{v}_{x}^{(1)}=-\frac{e}{m}\left(\widetilde{v}_y^{(1)}B_0\right)
\label{O_1st_diff}
\end{eqnarray}


\noindent
The solutions of Eq. \ref{O_1st_diff} can be written as

\begin{equation}
\widetilde{v}_{x}^{(1)}=\frac{e \omega_c}{m(\omega_c^2-\omega_{l}^2)}\widetilde{E}_l
\hspace{0.5cm};\hspace{0.5cm}
\widetilde{v}_{y}^{(1)}=\frac{i e \omega_{l}}{m(\omega_c^2-\omega_{l}^2)}\widetilde{E}_l
\hspace{0.5cm};\hspace{0.5cm}
\widetilde{v}_{z}^{(1)}=0
\label{O_mode_1st}
\end{equation}

\noindent
The second order terms of Eq. \ref{Lorntz} can be expressed as

\begin{equation}
\frac{\partial \widetilde{v}_{x}^{(2)}}{\partial t}=-\frac{e}{m}\left[\widetilde{v}_{y}^{(1)}\widetilde{B}_l+\widetilde{v}_{y}^{(2)}{B}_{0}\right]
\hspace{0.5cm};\hspace{0.5cm}
\frac{\partial \widetilde{v}_{y}^{(2)}}{\partial t}=\frac{e}{m}\left[\widetilde{v}_{x}^{(1)}\widetilde{B}_l+\widetilde{v}_{x}^{(2)}{B}_{0}\right]
\hspace{0.5cm};\hspace{0.5cm}
\frac{\partial \widetilde{v}_{z}^{(2)}}{\partial t}=0
\label{2nd_1st_diff}
\end{equation}

\noindent
Eq. \ref{2nd_1st_diff} can be expressed as the second order differential equations as follows:

\begin{eqnarray}
\frac{\partial^2 \widetilde{v}_{x}^{(2)}}{\partial t^2} + \omega_c^2\widetilde{v}_{x}^{(2)}=-\frac{e}{m}\left[\frac{\partial \left(\widetilde{v}_y^{(1)}\widetilde{B}_l\right)}{\partial t}+\omega_c\widetilde{v}_{x}^{(1)}\widetilde{B}_l\right]
\label{2nd_diff_x}
\end{eqnarray}

and, 

\begin{eqnarray}
\frac{\partial^2 \widetilde{v}_{y}^{(2)}}{\partial t^2} + \omega_c^2\widetilde{v}_{y}^{(2)}=\frac{e}{m}\left[\frac{\partial \left(\widetilde{v}_x^{(1)}\widetilde{B}_l\right)}{\partial t}-\omega_c\widetilde{v}_{y}^{(1)}\widetilde{B}_l\right]
\label{2nd_diff_y}
\end{eqnarray}

\noindent
We can solve the Eq. \ref{2nd_diff_x} and \ref{2nd_diff_y} using the expressions in Eq. \ref{O_mode_1st}. The solutions can be expressed as follows:

\begin{eqnarray}
\widetilde{v}_{x}^{(2)}=F_{1}exp(-i 2\omega_{l}t)
\hspace{0.5cm};\hspace{0.5cm}
\widetilde{v}_{y}^{(2)}=F_{2}exp(-i 2\omega_{l}t)
\label{O_mode_2nd}
\end{eqnarray}

Here, 

\begin{eqnarray}
F_1 = -\frac{(e/m)^2 E_{ly}^2\left(\omega_{c}^2 + 2\omega_{l}^2\right)}{c\left(\omega_c^2-4\omega_{l}^2\right)\left(\omega_c^2-\omega_{l}^2\right)}
\hspace{0.5cm};\hspace{0.5cm}
F_2 = -i\frac{3(e/m)^2 E_{ly}^2\omega_{c}\omega_{l}}{c\left(\omega_c^2-4\omega_{l}^2\right)\left(\omega_c^2-\omega_{l}^2\right)}
\end{eqnarray}

\noindent
From Eq. \ref{O_mode_2nd} we can obtain the $x$ and $y$ components of current oscillating with the frequency $2\omega_{l}$ (second harmonic) and is given by,

\begin{eqnarray}
\widetilde{J}_{ex}^{(2)}=-n_eeF_{1}exp(-i 2\omega_{l}t)
\hspace{0.5cm};\hspace{0.5cm}
\widetilde{J}_{ey}^{(2)}=-n_eeF_{2}exp(-i 2\omega_{l}t)
\label{O_mode_2nd_current}
\end{eqnarray}

\noindent
Similarly, the third order terms of Eq. \ref{Lorntz} can be expressed as

\begin{equation}
\frac{\partial \widetilde{v}_{x}^{(3)}}{\partial t}=-\frac{e}{m}\left[\widetilde{v}_{y}^{(2)}\widetilde{B}_l+\widetilde{v}_{y}^{(3)}{B}_{0}\right]
\hspace{0.5cm};\hspace{0.5cm}
\frac{\partial \widetilde{v}_{y}^{(3)}}{\partial t}=\frac{e}{m}\left[\widetilde{v}_{x}^{(2)}\widetilde{B}_l+\widetilde{v}_{x}^{(3)}{B}_{0}\right]
\hspace{0.5cm};\hspace{0.5cm}
\frac{\partial \widetilde{v}_{z}^{(3)}}{\partial t}=0
\label{3rd_1st_diff}
\end{equation}

From Eq. \ref{3rd_1st_diff}, we can write

\begin{eqnarray}
\frac{\partial^2 \widetilde{v}_{x}^{(3)}}{\partial t^2} + \omega_c^2\widetilde{v}_{x}^{(3)}=-\frac{e}{m}\left[\frac{\partial \left(\widetilde{v}_y^{(2)}\widetilde{B}_l\right)}{\partial t}+\omega_c\widetilde{v}_{x}^{(2)}\widetilde{B}_l\right]
\label{3rd_diff_x}
\end{eqnarray}

and, 

\begin{eqnarray}
\frac{\partial^2 \widetilde{v}_{y}^{(3)}}{\partial t^2} + \omega_c^2\widetilde{v}_{y}^{(3)}=\frac{e}{m}\left[\frac{\partial \left(\widetilde{v}_x^{(2)}\widetilde{B}_l\right)}{\partial t}-\omega_c\widetilde{v}_{y}^{(2)}\widetilde{B}_l\right]
\label{3rd_diff_y}
\end{eqnarray}

\noindent
Solving Eq. \ref{3rd_diff_x} and Eq. \ref{3rd_diff_y} by using the expression of $\widetilde{v}_{x}^{(2)}$ and $\widetilde{v}_{y}^{(2)}$ given in Eq. \ref{O_mode_2nd}, we can get

\begin{eqnarray}
\widetilde{J}_{ex}^{(3)}=-n_eeF_{3}exp(-i 3\omega_{l}t)
\hspace{0.5cm};\hspace{0.5cm}
\widetilde{J}_{ey}^{(3)}=-n_eeF_{4}exp(-i 3\omega_{l}t)
\label{O_mode_3rd_current}
\end{eqnarray}

Here, the expressions of $F_3$ and $F_4$ can be shown as

\begin{eqnarray}
F_3 = -\frac{(e/m) E_{ly}}{c\left(\omega_c^2-9\omega_{l}^2\right)}\left[F_1\omega_c - 3iF_2\omega_{l} \right]
\hspace{0.1cm};\hspace{0.1cm}
F_4 = -\frac{(e/m) E_{ly}}{c\left(\omega_c^2-9\omega_{l}^2\right)}\left[F_2\omega_c + 3iF_1\omega_{l} \right]
\end{eqnarray}

\noindent
Eq. \ref{O_mode_3rd_current} represents the $x$ and $y$ component of currents oscillating with the frequency $3\omega_{l}$ (third harmonic). It is to be noted that, unlike O-mode configuration, in this case we can get all the higher harmonics to be in the x-y plane. This has also been observed in our simulations.


\bibliography{ref}
\end{document}